\documentclass[rnote,traditabstract]{aa} % for the research notes
\usepackage{graphicx}
\usepackage{natbib}
\usepackage{amssymb}

\usepackage{txfonts}
\bibpunct{(}{)}{;}{a}{}{,}

\begin{document}
   \title{Polarization and photometric observations of the gamma-ray blazar
     PG\,1553+113\thanks{Based on observations collected at the Centro
       Astron\'omico Hispano Alem\'an (CAHA) at Calar Alto, operated jointly by
       the Max-Planck Institut f\"ur Astronomie and the Instituto de
       Astrof\'{\i}sica de Andaluc\'{\i}a (CSIC).}}

   \author{I. Andruchow\inst{1,2}%
   \and J. A. Combi\inst{1,3}
   \and A. J. Mu\~noz-Arjonilla\inst{4}
   \and G. E. Romero\inst{1,3}
   \and S. A. Cellone\inst{1,2}
   \and J. Mart\'{\i}\,\inst{4}}

   \institute{Facultad de Ciencias Astron\'omicas y Geof\'{\i}sicas,
     Universidad Nacional de La Plata, Paseo del Bosque, B1900FWA La Plata,
     Argentina.%
\and IALP, CONICET-UNLP, CCT La Plata, Paseo del Bosque, B1900FWA - La Plata - 
Argentina.%
\and IAR, CONICET, CCT La Plata, C.C. No. 5 (1894) Villa Elisa - Buenos Aires -
Argentina.%
\and Departamento de F\'{\i}sica (EPS), Universidad de Ja\'en, Campus Las  
Lagunillas s/n, A3, 23071 - Ja\'en - Spain.}

%   \date{Received ; accepted }

% \abstract{}{}{}{}{} 
% 5 {} token are mandatory
 
\abstract%
% context heading (optional)
% {} leave it empty if necessary  
{We present the results of an observational photo-polarimetry campaign of the 
blazar PG\,1553+113 at optical wavelengths. The blazar was recently detected
at very high energies ($> 100$\,GeV) by the H.E.S.S and MAGIC $\gamma$-ray
Cherenkov telescopes.

Our high-temporal resolution data show significant variations in the linear
polarization percentage and position angle at inter-night time-scales, while
at shorter (intra-night) time-scales both parameters varied less
significantly, if at all. Changes in the polarization angle seem to be
common in $\gamma$-ray emitting blazars.  Simultaneous differential
photometry (through the $B$ and $R$ bands) shows no significant variability
in the total optical flux.  We provide $B$ and $R$ magnitudes, along with a
finding chart, for a set of field stars suitable for differential
photometry.}


\maketitle

\section{Introduction}

The active galactic nucleus (AGN) \object{PG\,1553$+$113} was first
detected in the Palomar-Green survey of UV bright objects \citep{GSL86}.  It
is classified as a BL\,Lac object based on its featureless spectrum
\citep{MG83} and significant optical variability \citep{MCGH88}. The object
has been well studied from radio to VHE $\gamma$-ray ($>$ 100 GeV),
including several simultaneous multi-wavelength observation campaigns
\citep[see][]{O06, MDP09}.

At radio frequencies between 4.8\,GHz and 14.5\,GHz, the source was found
to be variable on timescales of months \citep{PMG05, O06}. VLBA observations
have shown a jet extending at least 20\,pc towards the northeast of the
object \citep{RGS03}.

In the X-ray band, PG\,1553$+$113 has been detected at different flux
levels by Einstein, ROSAT, ASCA \citep{DGTF01}, \emph{BeppoSAX}
\citep{DSG05}, RXTE \citep{O06}, \emph{XMM-Newton} \citep{PMG05}, Swift
\citep{TGM07}, and Suzaku \citep{RCM08} satellites. A list containing the 
flux and spectral parameters from all X-ray observations of the source 
can be found in \citet{AAA10}.

At very high energy (VHE) the object was first detected with the HESS
stereoscopic array of imaging atmospheric-Cherenkov telescopes during 2005
\citep{Ah06}. This detection was later
confirmed through MAGIC observations in 2005 and 2006 \citep{Al07}.
Both VHE spectra are unusually soft, although these results should be 
taken with caution because the errors are large.  
PG\,1553$+$113 was also detected 
at GeV $\gamma$-ray by the Fermi Gamma-ray Space Telescope \citep{Ab10}.

In this frequency range the source presented one of the hardest spectra
among the 106 AGNs listed in \citet{AAA09}. Combining archival radio, optical,
X-ray, and VHE $\gamma$-ray data, these authors modelled its spectral energy
distribution and demonstrated that a simple, one-zone synchrotron
self-Compton (SSC) model provides a reasonably good fit of the observational
results.  Unfortunately, the redshift of the source is still unknown. At
present no emission or absorption lines have been measured despite several
observational campaigns with optical instruments. However, an upper limit of
$z < 0.74$ was determined by \citet{Ah06} from the photon spectrum obtained
by HESS and later confirmed from MAGIC spectral measurements \citep{Al07}.

In this paper, we report optical polarimetric observations of the
$\gamma$-ray blazar PG\,1553$+$113. Our high-temporal resolution data show
significant variations in the position angle at intra-night as well as at
inter-night time-scales. Simultaneous differential photometry (at the $B$
and $R$ bands) shows no significant variability in the total optical flux.
We describe the observations and data analysis in
Sect.~\ref{Andruchow-Obs}. Then, we present the obtained results in
Sect.~\ref{Andruchow-Resu}, and we discuss the origin of the polarization
angle variations and state our conclusions in Sect.~\ref{Andruchow-Discu}.

\section{Observations and data reduction}
\label{Andruchow-Obs}

We observed PG\,1553+113 during six nights in April 2009 with the Calar
Alto Faint Object Spectrograph (CAFOS) in imaging polarimetry mode, at the
CAHA 2.2\,m. telescope, Calar Alto, Spain. The detector was a $2048 \times
2048$\,px blue sensitive SITe CCD, with 24\,$\mu$m pixels (the resulting
scale is $0.53$\,arcsec\,px$^{-1}$). In this mode, CAFOS is equipped with a
Wollaston prism plus a rotatable half-wave plate, which produce two
orthogonally polarized images --i.e., corresponding to the ordinary (O) and
the extraordinary (E) beams-- on the focal plane, with an effective beam
separation of $\sim 20$\,arcsec.  A mask with alternating blind and clear stripes
is placed before the detector to avoid image overlapping. Four
frames, each with a different position of the half-wave plate ($0$, $22.5$,
$45$, and $67.5$\,deg) are needed to obtain the normalised Stokes
parameters ($U$, $Q$) for linear polarization \citep[for more details, see][and
references therein]{CELL07}.

We observed through $B$ and $R$ Johnson-Cousins filters. The images range in
exposure time from 150 to 240\,s and from $50$ to $120$\,s for the $B$ and
$R$ filters, respectively.  Photometric data are obtained by adding up the
fluxes of the O and E images. In this way were able to follow simultaneously the
polarimetric and the photometric behaviour of the source.  Standard stars
from \citet{T90} were observed to determine the zero point for the
polarization angle and the instrumental polarization. The absolute
calibration of the photometry was performed by observing several standard
stars from \citet{L92} over a wide range of air masses during photometric
nights. For the same purpose, several images of the PG\,1553+113 field were
taken without the polarizing unit and through the $B$ and $R$ filters. The
images were debiased and flat-fielded in the standard way using the
IRAF\footnote{IRAF is distributed by the National Optical Astronomy
  Observatories, which are operated by the Association of Universities for
  Research in Astronomy, Inc., under cooperative agreement with the National
  Science Foundation.} reduction package.
Regrettably, the whole first night and part of the last two nights were
cloudy or with poor seeing conditions. We thus present results for the five
clear/useful nights from April 21 to April 25, 2009.

Instrumental magnitudes corresponding to the O and E images were obtained by
means of aperture photometry using the IRAF task \textsc{apphot}.  The same
process was performed for nine field stars suitably placed with respect to
the mask. We used these field stars to estimate the foreground polarization,
while two of them were selected to perform the differential
photometry (see \citealp{CRA07} for prescriptions on the appropriate choice of stars 
for differential photometry). In all the cases, a $3$~arcsec
aperture was used.

We obtained the linear polarization ($P$) and the position angle ($\theta$)
through each filter for PG\,1553+113 and the field stars along each night.
The data were corrected for instrumental polarization using the results for
the unpolarized standard stars.  The foreground polarization (formally, a
lower limit) was estimated from a suitable field star (\#9, see
Fig.\,\ref{findchart}), lying at 156\,arcsec projected distance from
PG\,1553+113, resulting in $P_\mathrm{fg}=0.24\%$. This value agrees 
with the low Galactic extinction in the direction to the blazar
($E(B-V)=0.052$~mag. $\Rightarrow P\lesssim 9 \times E(B-V)=0.468\%$,
\citealt{SFD98}). As a check, we measured the degree of polarization of
several other field stars. The mean value in both filters, $B$ and $R$, is
$< 1\%$, which confirms a low foreground polarization. The position angle
was transformed to the standard system using the highly polarized standard
stars.

\section{Results}
\label{Andruchow-Resu}

\subsection{Photometry}

\begin{figure}
\centering
\includegraphics[width=\hsize]{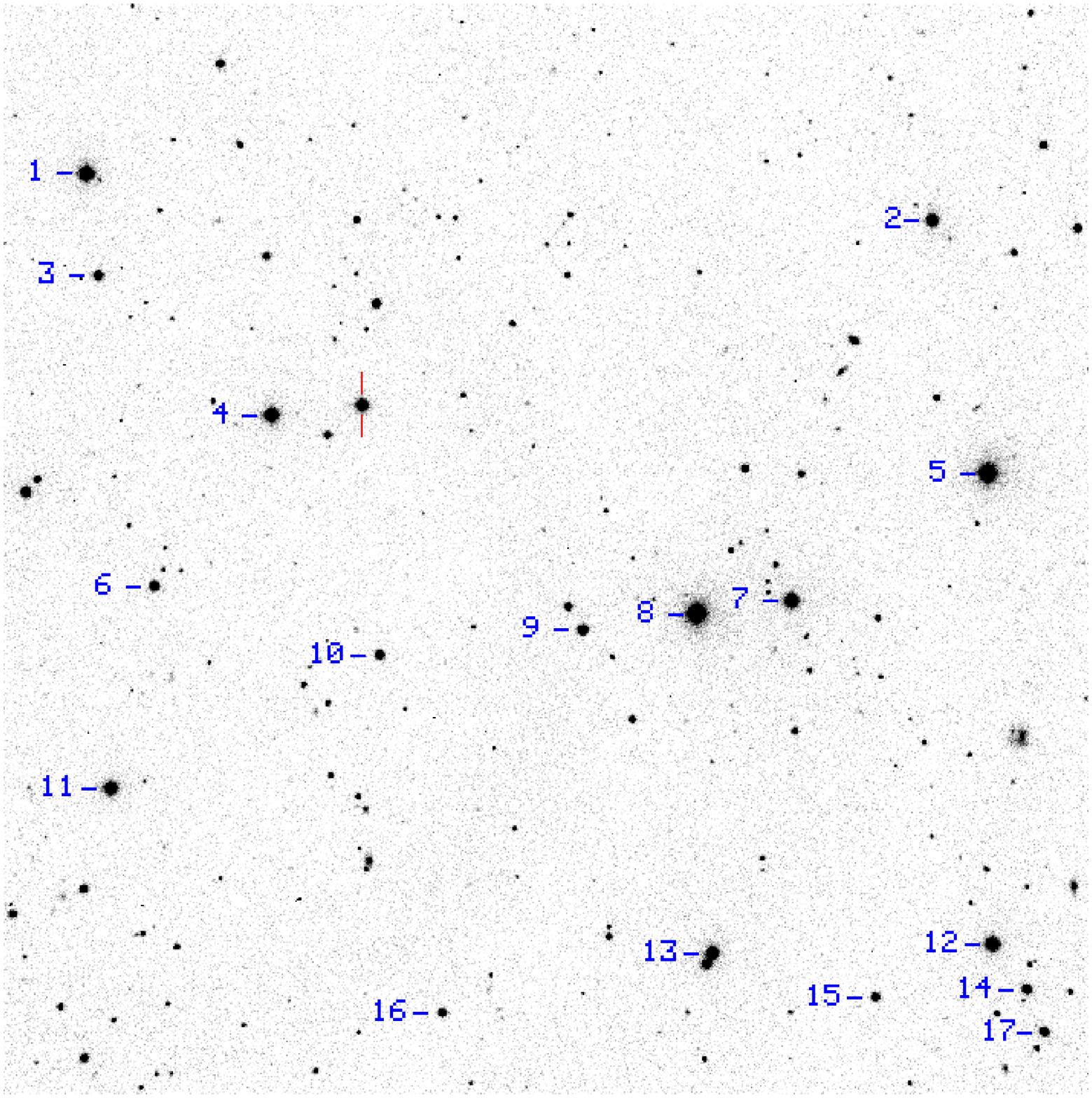}
\caption{Finding chart for comparison stars in the field of
  PG\,1553+113. The image is 9 arcmin on a side, with north up and east to
  the left. The blazar is bracketed by two vertical bars.\label{findchart}}
\end{figure}

The differential light-curves were obtained in the usual way, using a
non-variable star in the field as comparison, while another (also non-variable)
star was used to construct a second differential light-curve against the
comparison star, to be used for control purposes. Fig.\,\ref{findchart} is a finding 
chart for non-variable stars \footnote{The stability of these stars was verified 
with the 2007 data, see below.} suitable for differential photometry in the field of 
PG\,1553+113. Their magnitudes in the standard system \citep{L92} are given 
in Table~\ref{tab.std}. Throughout this study we used stars $12$ and $11$ 
as comparison and control, respectively.

\begin{table}
\caption{Standard magnitudes for field stars.} 
\label{tab.std}
\begin{tabular}{rcccc}
\hline\hline
 star  &  $B$  &  $\epsilon{B}$  &  $R$  &  $\epsilon{R}$ \\
\#   &  mag & mag & mag &  mag \\
\hline
  1   & 14.52  & 0.02 & 13.45 &  0.02 \\
  2   & 15.62  & 0.02 & 14.16 &  0.02 \\
  3   & 16.45  & 0.03 & 15.26 &  0.02 \\
  4   & 14.58  & 0.02 & 13.59 &  0.02 \\
  5   & 13.95  & 0.03 & 12.74 &  0.03 \\
  6   & 16.52  & 0.03 & 15.18 &  0.02 \\
  7   & 14.66  & 0.02 & 13.58 &  0.02 \\
  8   & 13.53  & 0.02 & 12.50 &  0.02 \\
  9   & 15.92  & 0.02 & 14.98 &  0.02 \\
 10   & 16.62  & 0.02 & 15.33 &  0.02 \\
 11   & 15.75  & 0.02 & 13.79 &  0.02 \\
 12   & 14.71  & 0.02 & 13.76 &  0.02 \\
 13   & 15.20  & 0.02 & 14.00 &  0.02 \\
 14   & 16.68  & 0.03 & 15.38 &  0.02 \\
 15   & 16.84  & 0.03 & 15.77 &  0.03 \\
 16   & 16.80  & 0.03 & 15.95 &  0.03 \\
 17   & 16.62  & 0.03 & 15.58 &  0.03 \\
\hline
\end{tabular}
\end{table}

A statistical analysis of these light-curves reveals that this VHE blazar
showed no significant variability in $B$ or in $R$. This is true
considering each individual night (microvariability) as well as taking the
five nights together (inter-night variability). The statistical 
analysis was performed as in \citet{CRA07} and applying the corrective
factor $\Gamma$ \citep[see][]{H88}, which weights the statistics
taking into account the differences in magnitude among the target and the
comparison and control stars. The values for $\Gamma$ were $\sim 0.75$
for $B$ and $\sim1.12$ for $R$\footnote{$\Gamma=1$ for the ideal case
when target and both stars all have the same magnitude.}.

The dispersions of the curves ranged from $0.004$ to $0.013$\,mag, and the 
mean apparent magnitudes of PG\,1553+113 for the whole campaign were 
$\langle B\rangle=14.90 \pm 0.003$\,mag and 
$\langle R \rangle=14.18 \pm 0.003$\,mag. Differential photometry data taken
with the 2.15-m ``Jorge Sahade'' telescope, at CASLEO, San Juan, Argentina,
during April 2007 seem to point in the same direction: the source was
non-variable at optical wavelengths. Because of the mask used for CAHA
observations, different comparison and control stars were used for both
data sets, but we were able to put all results on the same magnitude
scale because they had several field stars in common. Although the blazar's flux
remained steady during each of both periods, the object was about $0.1$ mag 
fainter in the $R$-band during 2009 compared with
2007. The results on PG\,1553+113 from the April 2007 observing run, along with 
those for other TeV $\gamma$-ray emitting blazars, will be presented
elsewhere (Andruchow et al., in preparation).

Both observations were performed after the high-energy detections 
(\citealt{Ah06} and \citealt{Al07}). On the other hand, the 
observations reported here were almost simultaneous with the final part 
of the first $\sim 200$ days of the \emph{Fermi Gamma-ray Space Telescope} 
science operations (ended on February 22, 2009), when PG\,1553+113 was 
studied in detail at the GeV gamma-ray regime \citep{Ab10}. The GeV 
emission state of PG\,1553+113 did not show any significant changes; 
this can be taken as an indication that it was in a low-energy state 
throughout the observations.

The photometric spectral index taking the five nights together and using the
mean magnitudes \citep[corrected for Galactic extinction,][]{SFD98} 
was $\alpha_{BR} = -0.71 \pm 0.01$. \citet{FSB94} reported a mean value for the
spectral index at visual wavelengths of $\langle \alpha \rangle= -0.96$,
obtained from continuum fitting to the optical spectrum.  Part of the difference
between the two values could arise because our $\alpha$ was
calculated using photometric observations, which could be affected by
atmospherics effects not completely removed (e.g., telluric lines). During
the period of five years in the early '90s, the authors reported that the
maximum variation in $\alpha$ was of about $0.24$.

\subsection{Polarimetry}

\begin{figure}
\includegraphics[width=0.95\hsize]{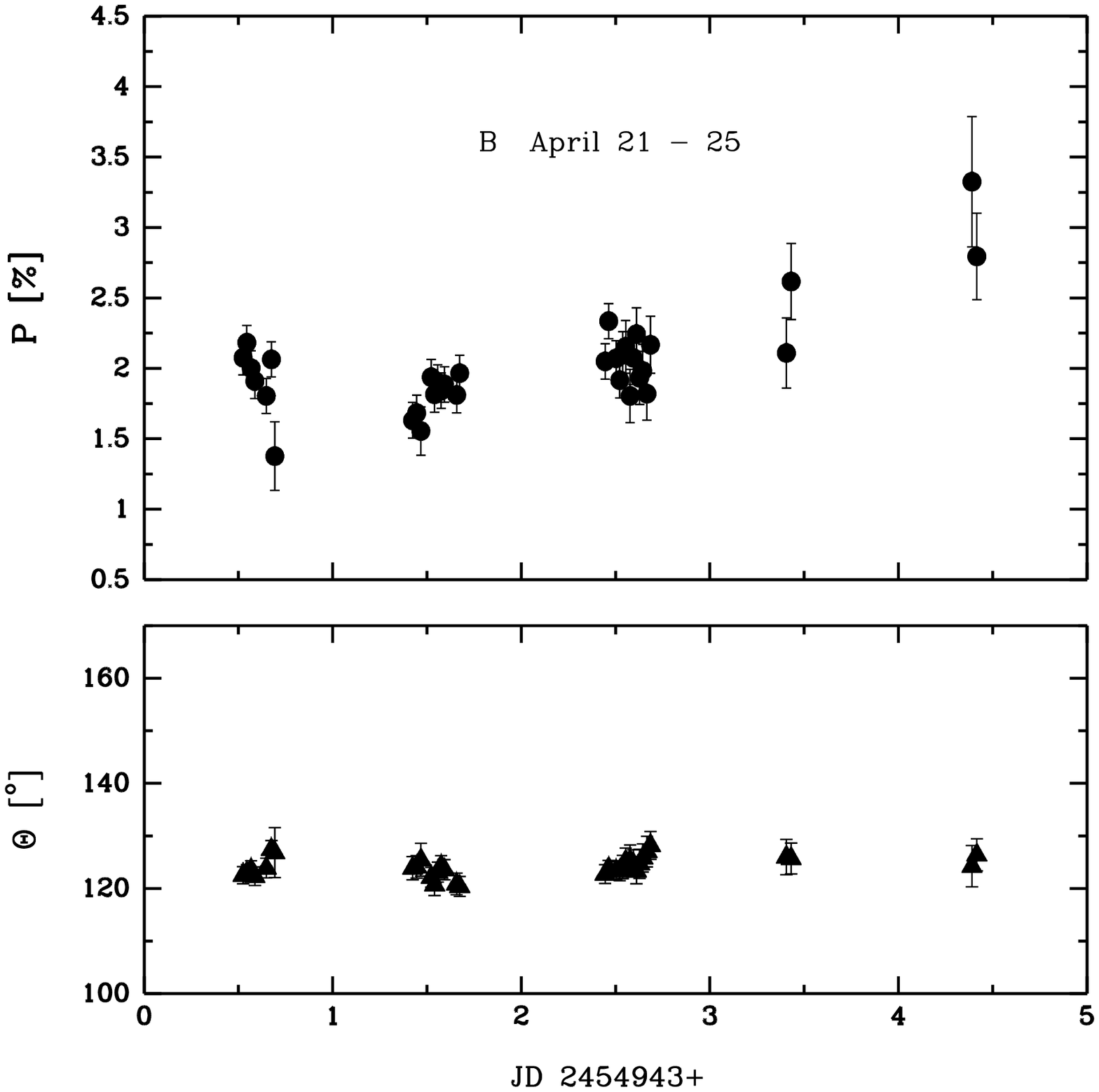} \\
\includegraphics[width=0.95\hsize]{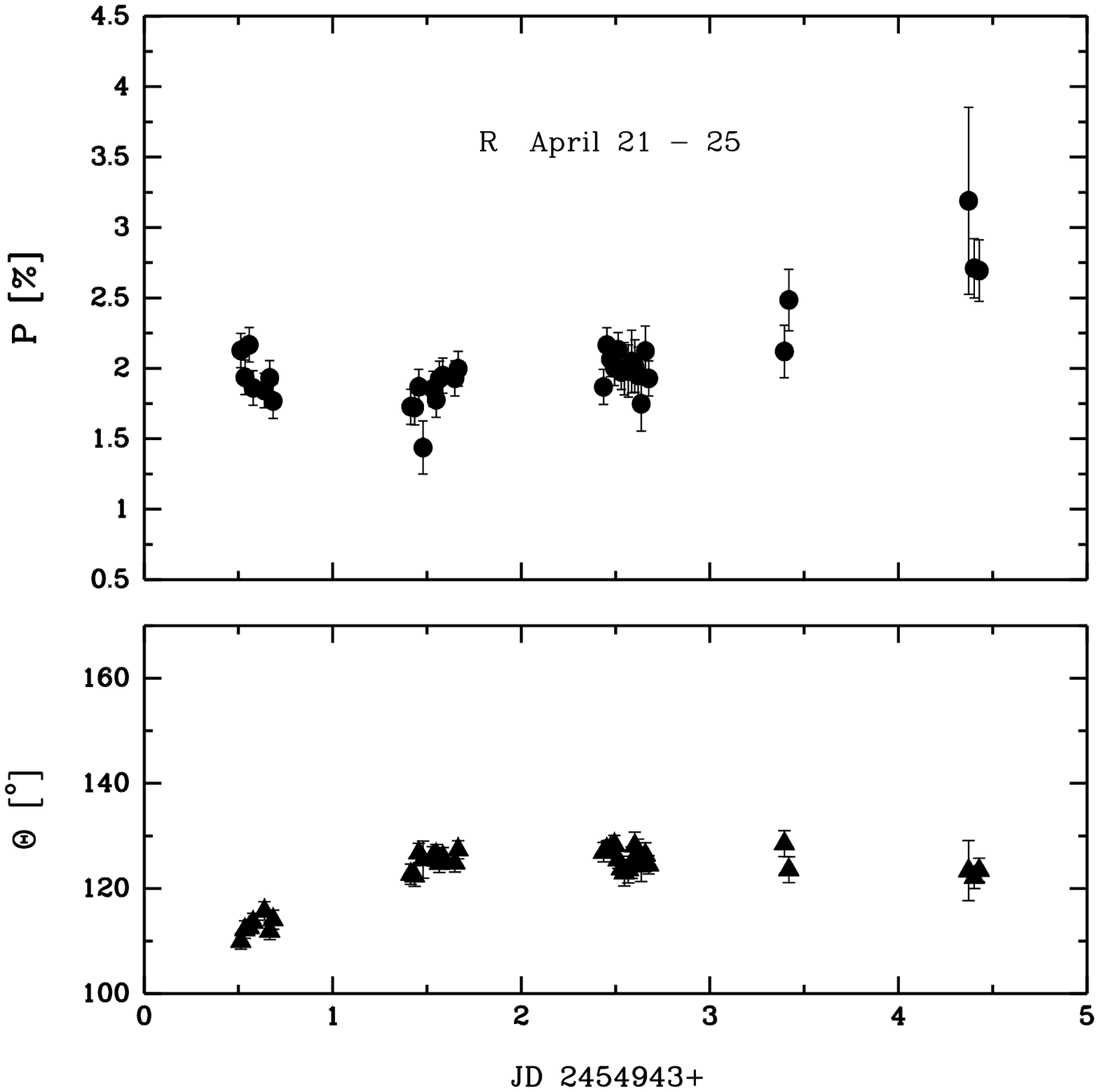}
\caption{Linear polarization (\textit{upper panels}) and position angle
  (\textit{lower panels}) against time for PG\,1553+113, through
  the $B$ filter (\textit{left}), and through the $R$ filter
  (\textit{right}).}
\label{pol}
\end{figure}

Figure~\ref{pol} presents the curves for the linear polarization and
position angle against time along the five nights.

Following a $\chi^{2}$ criterion, we performed the statistical analysis
for each night, in each filter, and also for the whole campaign. According
to this criterion, a source can be classified as variable in an observing
session for the observable $S$ if the probability of exceeding the value
$X^{2}$ by chance is $< 0.1$ \%, and it is classified as non-variable if the
probability is $> 0.5$ \%. If no definite decision could be
reached with that criterion (any value between $0.1$\% and $0.5$\%), we call 
this a \emph{dubious} behaviour. If the errors are random,
$X^2$ should be distributed as $\chi^2$ with $n-1$ degrees of freedom, where
$n$ is the number of points in the distribution. In Table \ref{polariza}
we show the values of the variability parameters for the
linear polarization percentage (upper section) and for the position angle (lower section).
Column 1 gives the observation date, Col.~2 gives the corresponding filter, 
Col.~3 shows the number of points for each night, 
Col.~4 gives the mean polarization (position angle) for the observing night,
Col.~5 shows the rms $\sigma_P$ ($\sigma_\theta$), Col.~6 gives the value
for $\Delta P$ ($\Delta \theta$) (the difference between the maximum and the
minimum values during the period reported), Col.~7 is $\Delta t$ (between
the maximum and the minimum), and Col.~8 shows the value of $\chi^2$.
 
\citet{A08} present a method to estimate the degree of polarization of the
nucleus, corrected for the de-polarizing effect from the host galaxy
light. This requires a detailed knowledge of the host's absolute magnitude
and effective radius, but this information is not available for PG\,1553+113
(its host remains undetected). However, from the upper limits on the
magnitudes given in \citet{USO00}, we do not expect that the host galaxy of
PG\,1553+113 would significantly reduce the degree of polarization of the
active nucleus, nor would it introduce any spurious variation.

\section{Discussion}
\label{Andruchow-Discu}

We found that the flux in both filters, $B$ and $R$, remained steady during
the observing run reported here. On the other hand, for the degree of
polarization and the position angle in the same period and in the same
filters, the source showed small variations. Comparing our results with
previous ones, we remark that \citet{O06} found in their multiwavelength
campaign (April-May 2003) that PG\,1553+113 was brighter in both optical
bands by about $0.5$~mag without signs of variation. Those observations
were made during an X-ray flare. In the same period, \citet{A05} followed
the source in linear optical polarization ($V$ filter) during two nights
with a good sampling in both nights. The degree of polarization and the
position angle then showed microvariations with a most likely flickering
behaviour. For the values reported here (see Table \ref{polariza}), we
note that the degree of polarization was higher during 2003, as was the
position angle. The polarization vector seems to have rotated, because the
position angle varied by $\sim20^{\circ}$ between both epochs, but it was still 
roughly orthogonal to the radio-jet direction reported by \citet{RGS03}.

This is particularly interesting given the recent detection of rapid optical
polarization changes correlated with $\gamma$-ray variability in the blazar
3C\,279 \citep{Fermi10}, which have been interpreted as the result of
geometrical effects \citep[see also][]{ACR03}.

We also have preliminary results from photometric and polarimetric data
taken in April 2007 \citep{A07} trough $R$ and $V$ filters. These show that
PG\,1553+113 presented no flux variations during the four nights during
which it was followed, although the blazar was then slightly brighter ($\sim
0.1$\,mag) than in 2009. Regarding the optical linear polarization, we 
obtained one point per night, with no significant variations. The mean 
value was $\langle P \rangle = 2.77\%$.

During 2005 and 2006 this source was detected at very high energies by HESS
and MAGIC (\citealt{Ah06} and \citealt{Al07}, respectively). This very high
emission was detected in between the optical observations that we discused above.

The data presented here can be useful to complete the multiwavelength view
of this peculiar object. In the near future, using deep exposures with the
Gemini North 8\,m telescope, we will attempt to determine the redshift
of the host galaxy.

\begin{table}
\caption{Statistical results for the linear polarization.} 
\label{polstat}\centering
\begin{tabular}{@{}cccccccc@{}} \hline \noalign{\vskip
2pt} \hline \noalign{\smallskip}
Date & {\small Filter} & n & $\langle P\rangle$ & $\sigma_{P}$ & $\Delta P$ &
$\Delta t$ & $\chi^{2}$\\
(2009)  & &  & $\%$~   & $\%$~ & $\%$~  & hr~ &  \\
\noalign {\smallskip} \hline\noalign{\vskip 2pt}
21 Apr & $B$ & 7 & 1.92 & 0.27 & 0.80 & 3.56 & 13.96~(4) \\
21 Apr & $R$ & 7 & 1.95 & 0.15 & 0.40 & 3.05 & \phantom{1}8.78~(4) \\
22 Apr & $B$ & 10 & 1.80 & 0.14 & 0.41 & 4.94 & \phantom{1}8.77~(1) \\
22 Apr & $R$ & 10 & 1.82 & 0.16 & 0.56 & 4.47 & 10.30~(3) \\
23 Apr & $B$ & 13 & 2.05 & 0.16 & 0.53 & 2.72 & 11.87~(8) \\
23 Apr & $R$ & 14 & 2.00 & 0.11 & 0.42 & 4.37 & \phantom{1}6.99~(1) \\
24 Apr & $B$ & 2 & 2.36 & 0.36 & 0.51 & 0.60 & \phantom{1}1.91~(8) \\
24 Apr & $R$ & 2 & 2.30 & 0.26 & 0.37 & 0.57 & \phantom{1}1.66~(6)\\
25 Apr & $B$ & 2 & 3.06 & 0.38 & 0.53 & 0.62 & \phantom{1}1.07~(4) \\
25 Apr & $R$ & 3 & 2.86 & 0.28 & 0.50 & 1.35 & \phantom{1}1.38~(4) \\
\noalign{\smallskip} \hline \noalign{\smallskip}
{\small All } & $B$ & 34 & 2.03 & 0.35 & 1.95 & 88.73 & 80.35~(1) \\
{\small All } & $R$ & 36 & 2.03 & 0.32 & 1.75 & 69.44 & 77.61~(8) \\ 
\noalign{\vskip 2pt} \hline
\noalign{\smallskip} \hline \noalign{\smallskip}
Date & {\small Filter} & & $\langle \theta\rangle$ & $\sigma_{\theta}$ & $\Delta
\theta$ & $\Delta t$ &  $\chi^{2}$ \\
(2009) &  &  & $^{\circ}$~ & $^{\circ}$~ & $^{\circ}$~& hr~ &  \\
\noalign {\smallskip} \hline\noalign{\vskip 2pt}
21 Apr & $B$ & 7 & 120.24 & 2.06 & 5.11 & 2.11 & \phantom{1}7.06~(4) \\
21 Apr & $R$ & 7 & 116.82 & 1.82 & 5.76 & 3.00 & \phantom{1}7.52~(7) \\
22 Apr & $B$ & 10 & 131.75 & 1.74 & 5.09 & 4.97 & \phantom{1}5.86~(4) \\
22 Apr & $R$ & 10 & 129.27 & 1.65 & 4.98 & 5.54 & \phantom{1}6.99~(3) \\
23 Apr & $B$ & 13 & 129.86 & 1.58 & 5.44 & 5.77 & \phantom{1}5.76~(5) \\
23 Apr & $R$ & 14 & 129.75 & 1.80 & 5.38 & 1.25 & 10.19~(2) \\
24 Apr & $B$ & 2 & 128.77 & 0.20 & 0.28 & 0.60 & \phantom{1}0.00~(4) \\
24 Apr & $R$ & 2 & 132.89 & 0.54 & 0.76 & 0.57 & \phantom{1}0.04~(8) \\
25 Apr & $B$ & 2 & 124.19 & 2.51 & 3.55 & 0.62 & \phantom{1}0.53~(9) \\
25 Apr & $B$ & 3 & 126.98 & 0.76 & 1.37 & 0.63 & \phantom{1}0.22~(2) \\
\noalign{\smallskip} \hline \noalign{\smallskip}
{\small All } & $B$ & 34 & 128.04 & 4.69 & 15.88 & 26.07 & 217.18~(4)  \\
{\small All } & $R$ & 36 & 127.05 & 5.45 & 19.34 & 69.75 & 361.33~(1) \\ 
\noalign{\smallskip} \hline
\end{tabular}
\label{polariza}
\end{table}

%%%%%%%%%%%%%%%%%%%%%%%%%%%%%%%figure%%%%%%%%%%%%%%%%%%%%%%%%
%%%%%%%%%%%%%%%%%%%%%%%%%%%%%%%%%%%%%%%%%%%%%%%%%%%%%%%%%%%

\begin{acknowledgements}
The authors acknowledge support for different aspects of this work by grants 
AYA2010-21782-C03-03 from the Spanish government, Consejer\'{\i}a de Econom\'{\i}a, 
Innovaci\'on y Ciencia of Junta de Andaluc\'{\i}a as research group FQM-322, 
excellence fund FQM-5418, FEDER funds, and by grants PICT 960 and PICT 
2008-0627, from ANPCyT, Argentina. G.E.R. and J.A.C. were supported by grant 
PICT 07-00848 BID 1728/OC-AR (ANPCyT) and PIP 2010-0078 (CONICET), Argentina. Skillful
assistance by the CAHA staff during the observations is also warmly acknowledged.

\end{acknowledgements}

\end{document}